\documentclass[aps,prc,twocolumn,showpacs,superscriptaddress]{revtex4}
\usepackage[english]{babel}
\usepackage{graphics}
\usepackage{amsmath}
\usepackage{amsfonts}
\usepackage{amssymb}
\usepackage{graphicx}
\usepackage{xcolor}

\begin{document}

\title{Magnetic hyperfine structure of the ground-state doublet
in highly charged ions $^{229}$Th$^{89+,87+}$ and the
Bohr-Weisskopf effect}

\author{E.~V.~Tkalya}
\email{tkalya@srd.sinp.msu.ru}

\affiliation{Skobeltsyn Institute of Nuclear Physics Lomonosov
Moscow State University, Leninskie gory, Moscow 119991, Russia}

\affiliation{Nuclear Safety Institute of RAS, Bol'shaya Tulskaya
52, Moscow 115191, Russia}

\affiliation{National Research Nuclear University MEPhI, 115409,
Kashirskoe shosse 31, Moscow, Russia}

\author{A.~V.~Nikolaev}

\affiliation{Skobeltsyn Institute of Nuclear Physics Lomonosov
Moscow State University, Leninskie gory, Moscow 119991, Russia}

\date{\today}

\begin{abstract}
The magnetic hyperfine (MHF) structure of the $5/2^+$(0.0 eV)
ground state and the low-lying $3/2^+$(7.8 eV) isomeric state of
the $^{229}$Th nucleus in highly charged ions Th$^{89+}$ and
Th$^{87+}$ is calculated. The distribution of the nuclear
magnetization (the Bohr-Weisskopf effect) is accounted for in the
framework of the collective nuclear model with the wave functions
of the Nilsson model for the unpaired neutron. The deviations of
the MHF structure for the ground and isomeric states from their
values in the model of point-like nuclear magnetic dipole are
calculated. The influence of the mixing of the states with the
same quantum number $F$ on the energy of sublevels is studied.
Taking into account the mixing of states, the probabilities of the
transitions between the components of MHF structure are found.

\end{abstract}

\pacs{32.10.Fn, 27.90.+b, 31.15.ve}

\maketitle

\section{Introduction}
\label{sec:Introduction}

The low lying $3/2^+ (7.8\pm0.5$ eV) state \cite{Beck-07,Beck-R}
of the $^{229}$Th nucleus has been the subject of intense
experimental
\cite{Beck-07,Beck-R,Reich-90,Helmer-94,Irwin-97,Richardson-98,
Shaw-99,Utter-99,Kikunaga-05,Campbell-09,Campbell-11,Campbell-12,
Hehlen-13,Jeet-15,Yamaguchi-15,Wense-16} and theoretical research
\cite{Strizhov-91,Tkalya-92,Wycech-93,Tkalya-96,Dykhne-96,Karpeshin-98,
Dykhne-98_ME,Dykhne-98,Tkalya-00-JETPL,Tkalya-00-PRC,Tkalya-03,
Flambaum-06,Berengut-09,Litvinova-09,Tkalya-11,Kazakov-12,Beloy-14,Tkalya-15}
in the past decades. The interest is caused by new possibilities
which are emerging in the study of such unusual nuclear level. The
works on this problem are multidisciplinary, including fields of
science as diverse as nuclear physics, solid state physics, atomic
physics, optics and laser physics. Three excellent experiments,
Refs.~\cite{Helmer-94,Beck-07,Wense-16}, which gave us the
knowledge on the existence of the low-lying isomeric state and
provided with estimations of its energy, demonstrate the variety
of experimental methods needed for the characterization of this
state. The experimental technique gradually evolves from
traditional methods of nuclear spectroscopy
\cite{Reich-90,Helmer-94,Beck-07} to those used in low energy
physics: solid state physics, optics, etc.
 \cite{Hehlen-13,Jeet-15,Yamaguchi-15,Wense-16}.

In a number of theoretical publications authors have drawn
attention to exciting possibilities related with the existence of
the $3/2^+(7.8\pm0.5$ eV) state and its decay channels. Here one
can mention a refinement of some fundamental laws and symmetries
of Nature, for example, the CP-violation, the variation of the
fine structure constant
\cite{Flambaum-06,Berengut-09,Flambaum-09,Dzuba-10,Skripnikov-14},
the local Lorentz invariance and the Einstein equivalence
principle \cite{Flambaum-16}. Unusual for traditional nuclear
physics decay channels of the isomeric state -- the electron
bridge \cite{Strizhov-91,Porsev-10-PRA} and nuclear light
\cite{Tkalya-00-JETPL,Tkalya-00-PRC,Tkalya-03} -- probably imply
that the $3/2^+(7.8\pm0.5$ eV) state can be occupied by laser
emission through nuclear photo-excitation process
\cite{Tkalya-96,Tkalya-03} or inverse electron bridge
\cite{Tkalya-92-JETPL,Tkalya-96,Porsev-10-PRL}. An interesting
consequence of such excitations is the detection of the
$\alpha-$decay of the isomeric state \cite{Dykhne-96} with the
possibility of checking the exponentiality of the decay law of an
isolated metastable state at long times \cite{Dykhne-98}. Finally,
we should mention two important technological applications -- the
nuclear clock \cite{Peik-03,Rellergert-10,Campbell-12,Peik-15},
and the gamma-ray laser of the optical range
\cite{Tkalya-11,Tkalya-13}, both of which can lead to a
breakthrough in their fields.

In the present study we have carried out calculations and give
numerical estimations for the position of the sublevels of the
$5/2^+(0.0)$ and $3/2^+(7.8$ eV) states in the highly charged ions
$^{229}$Th$^{89+}$ and $^{229}$Th$^{87+}$. The Th$^{89+}$ ion has
one electron which occupies the $1s_{1/2}$ electron shell, while
the Th$^{87+}$ ion has three electrons on the $1s_{1/2}$ and
$2s_{1/2}$ electron shells, i.e. the $(1s_{1/2})^2(2s_{1/2})^1$
electron configuration. We have taken into account the
Bohr-Weisskopf effect and the mixing of the states, and calculated
the probabilities of the transitions between the sublevels.

Our calculations can be instrumental in experimental studies of
the magnetic hyperfine (MHF) interaction in highly ionized atoms
investigated in a storage ring of accelerator complex
\cite{Geissel-92,Klaf-94,Radon-97,Ma-15}.

\section{Magnetic hyperfine interaction}

A preliminary (and as a rule, overestimated) estimation of the MHF
interaction can be obtained by using the Fermi contact interaction
\cite{Abragam-61}.

The electron in the $1s_{1/2}$ state of the $^{229}$Th$^{89+}$ ion
or electron in the $2s_{1/2}$ state of the $^{229}$Th$^{87+}$ ion
result in a strong magnetic field at the center of the $^{229}$Th
nucleus \cite{Wycech-93,Karpeshin-98}. The value of this field is
given by the formula for the Fermi contact interaction (see in
\cite{Abragam-61})
\begin{equation}
{\bf{H}}=-\frac{16\pi}{3} \mu_B \frac{{\mbox{\boldmath
$\sigma$}}}{2} |\psi_{e}(0)|^2,
\label{eq:H}
\end{equation}
where $\mu_B=e/2m$ is the Bohr magneton, $e$ is the electron
charge, $m$ is the electron mass, ${\mbox{\boldmath{$\sigma$}}}$
are the Pauli matrixes, and $\psi_{e}(0)$ is the amplitude of the
electron wave function at the origin.

The interaction of the point magnetic moment of the ground state
($\mu_{gr}=0.45$) or the isomeric state ($\mu_{is}=-0.076$) of the
$^{229}$Th nucleus with the magnetic field (\ref{eq:H}) leads to a
splitting of the nuclear levels. The energy of the sublevels is
determined by the formula
\begin{equation}
E=E_{int}\frac{F(F+1)-I(I+1)-s(s+1)}{2Is},
\label{eq:E}
\end{equation}
with the interaction energy
\begin{equation}
E_{int}=-\mu_{gr(is)}\mu_N{}H . \label{eq:Eint_Fermi}
\end{equation}
Here $\mu_N=e/2M_p$ is the nuclear magneton
($M_p$ is the proton mass), $I$ stands for the nuclear spin, $s$ is
the electron spin. The quantum number $F$ takes two values
$F=I\pm1/2$ for both the ground and isomeric state.

To obtain the magnetic field and the hyperfine splittings one has
to find the amplitude of the electron wave function at the
nucleus, $\psi_{e}(0)$, Eq.\ (\ref{eq:H}). The details of the
calculations of $\psi_{e}(0)$ are given in the next section. For
precise determination of $\psi_{e}(0)$ the calculations should be
self-consistent, relativistic and take into account the finite
size of the nucleus.

\section{Calculation of electron wave functions}

In the present work, the wave function $\psi_{1s_{1/2}}(x)$ of the
electron in the $(1s_{1/2})^1$ configuration has been found
numerically by solving the Dirac equation in the Coulomb potential
of the nucleus with the radius $R_0=1.2\, A^{1/3}$~fm (where $A$
is the atomic number). The nucleus has been considered spherical
with the homogeneous positive charge distribution. In this case
the electron potential energy in the units of the electron mass is
given by
\begin{equation}
V(x)  = \left \{
\begin{array}{ll}
\frac{1}{2}\frac{e^2Z}{mR_0}(3-x^2)\  , & \quad x\leq1 \\
\frac{e^2Z}{mR_0}\frac{1}{x}\  , & \quad x>1 \quad .
\end{array} \right.
\label{eq:Potential}
\end{equation}
Here $x=r/R_0$, where $r$ is the modulus of the electron radius
vector ${\bf{r}}$, $Z$ is the nuclear charge ($Z=90$ for Th). The
large ($g$) and small ($f$) radial components of the electron wave
function $\psi_{e}(x)$ are found from the system
\begin{equation}
\begin{array}{l}
xg'(x)-b(E+1-V(x))xf(x)=0,\\
xf'(x)+2f(x)+b(E-1-V(x))xg(x)=0,
\end{array}
\label{eq:Dirac}
\end{equation}
where $b=mR_0$. The electron wave function is normalized by the
condition $\int_0^{\infty}(g^2(x)+f^2(x))x^2dx=1$.

In the following we work with the variable $x=r/R_0$ instead of
$x=r/a_B$, where $a_B$ is the Bohr radius, commonly used in atomic
physics because for calculations of the Bohr-Weisskopf effect we
need a nuclear scale. However, when necessary we will return to
atomic units and give the values of the electron wave functions in
units of $a_B^{3/2}$ to compare with the calculations of other
authors.

\begin{figure}
 \includegraphics[angle=0,width=0.8\hsize,keepaspectratio]{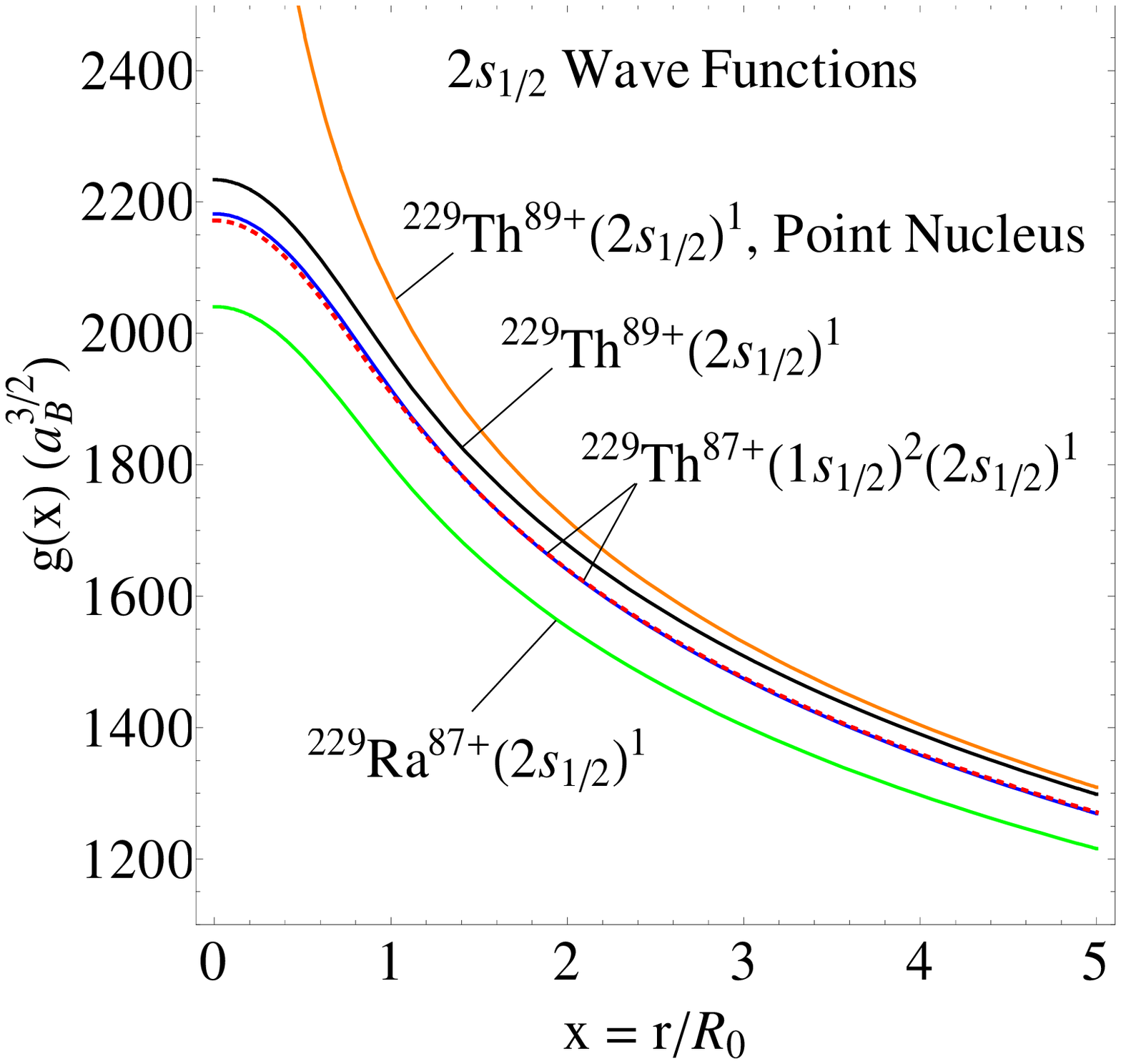}
  \caption{(color online).
The large radial component $g(x)$ of the $2s_{1/2}$ state (in
a.u.) calculated in various models (see text for details).
$^{229}$Th$^{89+}(2s_{1/2})^1$ is the hygrogen-like ion model for
the $2s_{1/2}$ state (i.e. the model without screening).
$^{229}$Ra$^{87+}(2s_{1/2})^1$ stands for the full screening of
the nuclear charge +2 by two electrons in the $1s_{1/2}$ state.
$^{229}$Th$^{87+}(1s_{1/2})^2(2s_{1/2})^1$, full line: the
potential from two electrons in the $1s_{1/2}$ state is added to
the nuclear potential. $^{229}$Th$^{87+}(1s_{1/2})^2(2s_{1/2})^1$,
dashed line: {\it ab initio} DFT calculation
\cite{Nikolaev-15,FLAPW} }
  \label{fig:2S_WaveFunctions}
\end{figure}

The calculation of the Bohr-Weisskopf effect for the $2s_{1/2}$
electron state in the $(1s_{1/2})^2(2s_{1/2})^1$ three electron
configuration has been performed by three different methods,
Fig.~\ref{fig:2S_WaveFunctions}.

%
%  Figure 1
%

At first, we have found the upper and lower boundaries for the
$2s_{1/2}$ electron wave function. For this purpose we have used
the wave function of the $2s_{1/2}$ state obtained in the field of
the hydrogen-like ion. As the hydrogen-like ion we take (i) the
$^{229}$Th$^{89+}$ ion with $Z=90$ and (ii) the $^{229}$Ra$^{87+}$
ion. In the last case we have assumed that two electrons in the
$1s_{1/2}$ state perfectly screen the nuclear charge +2, and
$Z=90-2=88$.

In the second model we have used a more sophisticated approach by
adding to the nuclear potential (\ref{eq:Potential}) the Coulomb
contribution from the electron density of two electrons in the
$1s_{1/2}$ state. The shortcoming of this calculation is that it
does not take into account the interaction between electrons in
the $1s_{1/2}$ state and between electrons in the $1s_{1/2}$ and
$2s_{1/2}$ states.

Finally, we have carried out {\it{ab initio}} numerical
calculations inside and outside the nuclear region within the
density functional method (DFT) taking the full account of the
electron self-consistent field by employing the atomic part of the
code \cite{Nikolaev-15,FLAPW}. The precise atomic DFT calculations
have been performed with two main variants for the exchange
correlation potential and energy. The first functional belongs to
the local density approximation (LDA) \cite{Dirac-30,Perdew-92},
while the second to the generalized gradient approximation (GGA)
of Perdew, Burke, and Ernzerhof (PBE) \cite{Perdew-96,Perdew-97}.
Both variants represent standard choices within the DFT method.

As follows from Fig.~\ref{fig:2S_WaveFunctions} the refined
calculation of the large component $g_{2s_{1/2}}(x)$ of the
$2s_{1/2}$ state practically coincides with the calculation in the
second model with the additional potential created by the two
noninteracting $1s_{1/2}$ electrons. Moreover, for some
estimations of the magnetic field and MHF interaction one can use
the simple calculation in the hygrogen-like ion model for
$^{229}$Th$^{89+}(2s_{1/2})^1$. The difference in the amplitude of
the $2s_{1/2}$ large component at nucleus, $g_{2s_{1/2}}(0)$, for
the $(2s_{1/2})^1$ configuration in $^{229}$Th$^{89+}$ and for the
$(1s_{1/2})^2(2s_{1/2})^1$ configuration in $^{229}$Th$^{87+}$
amounts to 2.5\%. For many applications this difference is not
essential.

To evaluate the magnetic field one can use Eq.~(\ref{eq:H}) with
$ \psi_{e}(0)=Y_{00}(\vartheta,\varphi)g(0)/R_0^{3/2} $,
where $Y_{00} = 1/sqrt{4\pi}$ is the spherical harmonic with
$l=m=0$ describing the angular part of the $s-$states. (Notice
that for $s-$states $f(x=0)=0$.) We have  $g_{1s_{1/2}}(0) =
8.46\times10^{-3}$ for the $(1s_{1/2})^1$ configuration in the
$^{229}$Th$^{89+}$ ion and $g_{2s_{1/2}}(0) = 3.57\times10^{-3}$
for the $2s_{1/2}$ state in the $(1s_{1/2})^2(2s_{1/2})^1$
configuration of the $^{229}$Th$^{87+}$ ion according to the
{\it{ab initio}} calculations. Now by means of Eq.~(\ref{eq:H}) we
obtain 112 MT for the magnetic field at the $^{229}$Th nucleus of
the $^{229}$Th$^{89+}$ ion and 20 MT for the magnetic field of the
$^{229}$Th$^{87+}$ ion.

Notice that in the atomic units
$ \psi_{e}(0)=Y_{00}(\vartheta,\varphi) \, g(0)/a_B^{3/2}$,
where $g_{1s_{1/2}}(0) = 5.18\times10^3$, and $g_{2s_{1/2}}(0) =
2.18\times10^3$.

\section{The Bohr-Weisskopf effect}

The influence of the finite nuclear size on the hyperfine
splitting was first considered by Bohr and Weisskopf
\cite{Bohr-50}. Later the effect of the distribution of nuclear
magnetization on hyperfine structure in muonic atoms was studied
by Le Bellac \cite{LeBellac-63}. According to their works,
the energy of sublevels of a deformed nucleus is given by
Eq.~(\ref{eq:E}), where
\begin{equation}
E_{int}=\int{}d^3r\,{\bf{j}}({\bf{r}}) {\bf{A}}({\bf{r}}),
\label{eq:Eint}
\end{equation}
is the energy of the interaction of the electron current
${\bf{j}}({\bf{r}}) = -e\psi_{e}^+({\bf{r}}){\mbox{\boldmath
$\alpha$}} \psi_{e}({\bf{r}})$ (${\mbox{\boldmath
$\alpha$}}=\gamma^0{\mbox{\boldmath $\gamma$}}$, $\gamma$ are the
Dirac matrices) with the vector potential of the electromagnetic
field ${\bf{A}}({\bf{r}})$ generated by the magnetic moment of the
nucleus.

For a system of ``rotating deformed core (with the collective
rotating angular momentum ${\mbox{\boldmath $\Re$}}$) + unpaired
neutron (with the spin ${\bf{S}}_n$)'', the vector-potential is
determined by the relation \cite{Bohr-50,LeBellac-63}

\begin{eqnarray}
&&{\bf{A}}({\bf{r}}) =
-\int{}d^3R\left[\rho_n({\bf{R}})g_S{\bf{S}}_n + \right. \nonumber\\
&&\qquad\left. \rho_{core}^m({\bf{R}})g_R{\mbox{\boldmath $\Re$}}
\right] \times {\mbox{\boldmath
$\nabla$}}_r\frac{1}{|{\bf{r}}-{\bf{R}}|},
 \label{eq:A}
\end{eqnarray}
where $\rho_n({\bf{R}})$ is the distribution of the spin part of
the nuclear moment and $\rho_{core}^m({\bf{R}})$ is the
distribution of the core magnetization, $g_S$ is the spin
$g$-factor, and $g_R$ is the core gyromagnetic ratio. The
distributions $\rho_n({\bf{R}})$ and $\rho_{core}^m({\bf{R}})$ are
normalized by the conditions $\int{}d^3R \rho_n({\bf{R}})=1$,
$\int{}d^3R \rho_{core}^m({\bf{R}})=1$.

As follows from Eqs.~(\ref{eq:Eint}--\ref{eq:A}), $E_{int}$
consists of two parts, $E_{int} = E_{int}^{(n)} +E_{int}^{(core)}
$. Here $E_{int}^{(n)}$ is the energy of the electron interacting
with an external unpaired neutron and $E_{int}^{(core)}$ is the
energy of the electron interacting with the rotating charged
nuclear core. These energies are calculated in accordance with
formulas from \cite{LeBellac-63}. In our case the electron
interacts with the nucleus in the head levels of rotational bands
(for such states we have $K=I$, where $K$ is the component of $I$
along the symmetry axis of the nucleus), and two contributions are
given by
\begin{eqnarray}
E_{int}^{(n)}&=&-\frac{2e^2M_p}{3(M_pR_0)^2}g_K
\frac{I^2}{I+1}\left[\int_0^{\infty}f(x)g(x)dx - \right.  \nonumber\\
&&\int{}\varphi^*_K({\bf{y}})\varphi_K({\bf{y}})d^3y
\int_0^y\left(1-x^3\frac{2I+1}{I(2I+3)}\times \right. \nonumber\\
&& \left. \left.
\sqrt{\frac{4\pi}{5}}Y_{20}(\theta)\right)f(x)g(x)dx\right],
 \label{eq:EintNeutron}
\end{eqnarray}
\begin{eqnarray}
E_{int}^{(core)}&=&-\frac{2e^2M_p}{3(M_pR_0)^2}g_R
\frac{I}{I+1}\left[\int_0^{\infty}f(x)g(x)dx - \right.  \nonumber\\
&&\left.\int{}\rho_{core}^m({\bf{y}})d^3y
\int_0^y(1-x^3)f(x)g(x)dx\right].
 \label{eq:EintCore}
\end{eqnarray}
Here, $\varphi_K({\bf{R}})$ is the wave function of the external
neutron (see below), $g_K$ is the intrinsic $g$ factor,
${\bf{y}}={\bf{R}}/R_0$, ${\bf{R}}$ is the radius vector of the
unpaired neutron, and $\rho_{core}^m({\bf{y}})$
is the normalized nuclear magnetic moment. In the present approach
we consider the homogeneous positive charge distribution inside the nuclear sphere,
which results in
$$
\rho_{core}^m({\bf{y}})=\frac{5}{2}\frac{1}{4\pi{}R_0^3/3}y^2
\sin^2\theta.
$$

We conclude this section by noting that the first term in the
square brackets of Eqs.~(\ref{eq:EintNeutron}) and
(\ref{eq:EintCore}), $\int_0^{\infty}f(x)g(x)dx$, corresponds to
the interaction of the electron with a point nuclear magnetic
dipole. The model of the electron interacting with the point
nuclear magnetic dipole gives much more precise value of the
hyperfine interaction than the Fermi contact interaction,
Eqs.~(\ref{eq:H}) and (\ref{eq:Eint_Fermi}).

\section{Nuclear wave functions}

For calculations of the nuclear part in
Eqs.~(\ref{eq:EintNeutron})--(\ref{eq:EintCore}) we use the
standard nuclear wave function \cite{Bohr-98-II}
\begin{equation}
\Psi_{MK}^I=\sqrt{\frac{2I+1}{8\pi^2}}D_{MK}^I({\bf{\Omega}})\varphi_K({\bf{R}}),
\end{equation}
where $D_{MK}^I({\bf{\Omega}})$ is the Wigner $D$-function
\cite{Varshalovich-88}, ${\bf{\Omega}}$ stands for the three Euler
angles, $\varphi_K({\bf{R}})$ is the wave function of the external
neutron coupled to the core, and $M$ is the component of $I$ along
the direction of magnetic field.

The wave functions $\varphi_K$ of the unpaired neutron is taken
from the Nilsson model. The structure of the $^{229}$Th ground
state $5/2^+(0.0)$ is $K^{\pi}[Nn_z\Lambda]=5/2^+[633]$. The
structure of the isomeric state $3/2^+$(7.8 eV) is $3/2^+[631]$
\cite{Helmer-94}. For each of these states, the wave function has
the form
\begin{equation}
\varphi_K
=\phi_{\Lambda}(\varphi)\phi_{\Lambda,n_r}(\eta)\phi_{n_z}(\zeta)
\end{equation}
where the quantum number $n_r=(N-n_z-\Lambda)/2$,
$\zeta=R_0\sqrt{M_p\omega_z}\, y\, \cos\theta$ and
$\eta=R_0\sqrt{M_p\omega_{\perp}}\, y\, \sin\theta$ are new
variables. Here we have introduced new frequencies
$\omega_z=\omega_0\sqrt{1+2\delta/3}$ and
$\omega_{\perp}=\omega_0\sqrt{1-4\delta/3}$, where
$\omega_0=41/A^{1/3}$ MeV is the harmonic oscillator frequency,
$\delta=0.95\beta$, and $\beta$ is the parameter of the nuclear
deformation defined by $R=R_0(1+\beta{}Y_{20}(\theta)+\ldots)$.

For the wave function components we then obtain:
\begin{eqnarray*}
\phi_{\Lambda}(\varphi) &=& \frac{1}{\sqrt{2\pi}} e^{i\Lambda\varphi},\\
\phi_{\Lambda,n_r}(\eta) &=& \frac{1}{N_{\eta}} e^{-\eta^2/2}
\eta^{\Lambda}
L_{n_r}^{(\Lambda)}(\eta^2),\\
\phi_{n_z}(\zeta) &=& \frac{1}{N_{\zeta}}
 e^{-\zeta^2/2}H_{n_z}(\zeta),
\end{eqnarray*}
where $L_{n_r}^{(\Lambda)}$ is the generalized Laguerre
polynomial, $H_{n_z}$ is the Hermite polynomial
\cite{Abramowitz-64}, $N_{\eta,\zeta}$ are the normalization
factors. In our numerical calculations we took into account the
asymmetry of the nucleon wave functions in
Eq.~(\ref{eq:EintNeutron}), but neglected the small difference
between $\omega_z$ and $\omega_{\perp}$.

\section{Magnetic hyperfine structure and mixing of sublevels}
\label{sec:MHFS}

The energies of the sublevels calculated according to Eqs.\
(\ref{eq:E}), (\ref{eq:EintNeutron})-(\ref{eq:EintCore}) are given
in Tables~\ref{tab:HPS_229Th89+} -- \ref{tab:HPS_229Th87+}, third
columns. We observe that for the $^{229}$Th$^{89+,87+}$ ions there
is a reduction of the MHF splitting in comparison with the model
of point nucleus. The reduction is 3\% for the $5/2^+(0.0)$ ground
state and approximately 6\% for the $3/2^+$(7.8 eV) isomeric
state. The difference can be explained by the following: The
magnetic field produced by the spin of the nucleon is sensitive to
the non-sphericity of the wave functions $\varphi_K$. This leads
to the appearance of the additional averaging over the angle
$\theta$ in Eq.~(\ref{eq:EintNeutron}) \cite{Bohr-50,LeBellac-63}.
Averaging over the angles reduces the spin contribution with
respect to the orbital part. A small imbalance emerged in the
system leads to some violation of the ``fine tuning'' between the
spin and orbital parts of the magnetic moment. The relative
imbalance is larger for the isomeric state because its magnetic
moment smaller than the magnetic moment of the ground state.

%
% Table 1
%
\begin{table}
  \caption{Magnetic hyperfine splitting in the $^{229}$Th$^{89+}$ ion.
  The energy of the sublevels is in eV.}
  \begin{tabular}{c|c|c|c|c}
    \hline
    \hline
                & \multicolumn{1}{c|}{Point}   &\multicolumn{2}{c|} {Distributed nuclear} \\
    State       & \multicolumn{1}{c|}{nuclear} &\multicolumn{2}{c|} {magnetic dipole}     \\
                 \cline{3-4}
    $I^{\pi},F$ & magnetic                     & Eqs.\ (\ref{eq:EintNeutron})-(\ref{eq:EintCore})    & Mixing of the \\
                & dipole                       & of the present work                                 & $F=2$ levels  \\
    \hline
    $5/2^+, 3$  &  0.373                       &    0.362                                            &  0.362        \\
    \hline
    $5/2^+, 2$  &  -0.522                      &    -0.507                                           &  -0.526       \\
    \hline
    $3/2^+, 2$  &  $E_{is}-0.063$              &  $E_{is}-0.059$                                     & $E_{is}-0.040$ \\
     \hline
    $3/2^+, 1$  &  $E_{is}+0.105$              &  $E_{is}+0.098$                                     & $E_{is}$+0.098 \\
    \hline
  \end{tabular}
  \label{tab:HPS_229Th89+}
\end{table}
%

%
% Table 2
%
\begin{table}
  \caption{Magnetic hyperfine splitting in the $^{229}$Th$^{87+}$ ion.
  The energy of the sublevels is in eV.}
  \begin{tabular}{c|c|c|c|c}
    \hline
    \hline
                & \multicolumn{1}{c|}{Point}   &\multicolumn{2}{c|} {Distributed nuclear} \\
    State       & \multicolumn{1}{c|}{nuclear} &\multicolumn{2}{c|} {magnetic dipole}     \\
                 \cline{3-4}
    $I^{\pi},F$ & magnetic                     & Eqs.\ (\ref{eq:EintNeutron})-(\ref{eq:EintCore})    & Mixing of the \\
                & dipole                       & of the present work                                 & $F=2$ levels  \\
    \hline
    $5/2^+, 3$  &  0.0616                      &    0.0597                                           &  0.0597        \\
    \hline
    $5/2^+, 2$  &  -0.0862                     &    -0.0836                                          &  -0.0841       \\
    \hline
    $3/2^+, 2$  &  $E_{is}-0.0104$             &  $E_{is}-0.0097$                                    & $E_{is}-0.0092$ \\
     \hline
    $3/2^+, 1$  &  $E_{is}+0.0173$             &  $E_{is}+0.0161$                                    & $E_{is}$+0.0161 \\
    \hline
  \end{tabular}
  \label{tab:HPS_229Th87+}
\end{table}

Since two sublevels in the ions $^{229}$Th$^{89+,\, 87+}$ has the
same quantum number $F$, one has to take into account the mixing
of these states \cite{Wycech-93,Karpeshin-98}:
\begin{equation}
\begin{array}{l}
|3/2^+,F=2\rangle{}' = \sqrt{1-b^2}|3/2^+,F=2\rangle + b|5/2^+,F=2\rangle \nonumber \\
|5/2^+,F=2\rangle{}' = \sqrt{1-b^2}|5/2^+,F=2\rangle -
b|3/2^+,F=2\rangle ,\nonumber
\end{array}
\end{equation}
where $b$ is the mixing coefficient given by
\cite{Davydov-65}
$$
b=\frac{E_{|3/2^+,F=2\rangle{}'}-E_{|3/2^+,F=2\rangle}}
{\sqrt{(E_{|3/2^+,F=2\rangle{}'}-E_{|3/2^+,F=2\rangle})^2+E_{M1}^2}} .
$$
It can be considered as a function of the energy $E_{M1}$ related
with the interaction of the electron and nuclear currents in the
$M1$ electron transition in the ionic shell and the $M1$ nuclear
transition between the ground and isomeric states. According to
\cite{Tkalya-92} the energy for the $E(M)L$ transition can be
found as
\begin{eqnarray}
E_{E(M)L}^2 &=& 4\pi{}e^2 \frac{\omega_N^{2(L+1)}}{[(2L+1)!!]^2}
\left(C_{j_i1/2L0}^{j_f1/2}\right)^2 \times \nonumber\\
&&\langle{}f|{\emph{m}}_L^{E(M)}|i\rangle^2
B(E(M)L,I_i\rightarrow{}I_f), \label{eq:E_L}
\end{eqnarray}
where $C_{j_i1/2L0}^{j_f1/2}$ stands for the Clebsh-Gordon
coefficients, $\langle{}f|{\emph{m}}_L^{E(M)}|i\rangle$ is the
electron matrix element and $B(E(M)L,I_i\rightarrow{}I_f)$ is the
reduced probability of the nuclear transition \cite{Bohr-98-I}.

In our case as a result of the $M1$ transition
$|1s_{1/2},3/2^+\rangle \rightarrow |1s_{1/2},5/2^+\rangle$ in the
$^{229}$Th$^{89+}$ ion or the $M1$ transition
$|2s_{1/2},3/2^+\rangle \rightarrow |2s_{1/2},5/2^+\rangle$ in the
$^{229}$Th$^{87+}$ ion, Eq.~(\ref{eq:E_L}) can be rewritten as
\begin{eqnarray}
E_{M1}^2 &=& \frac{15}{2}
\left(\frac{-2e^2M_p}{3(M_pR_0)^2}\right)^2
|\langle{}f|{\emph{m}}_1^{M}|i\rangle|^2 \times \nonumber\\
&& B_{W.u.}(M1,3/2^+\rightarrow{}5/2^+) .
\label{eq:Eint_M1}
\end{eqnarray}
Here $B_{W.u.}(M1,3/2^+\rightarrow{}5/2^+)$ is the reduced
probability of the nuclear isomeric transition in Weisskopf's
units \cite{Blatt-Weisskopf} and we have used that
$B_{W.u.}(M1,3/2^+\rightarrow{}5/2^+)=3.0\times 10^{-2}$
\cite{Tkalya-15}. The electron matrix element of the $M1$
transition \cite{Tkalya-92} reads as
\begin{eqnarray*}
\langle{}f|{\emph{m}}_1^{M}|i\rangle &=& (\kappa_i+\kappa_f)
\int_0^{\infty} h_1^{(1)}(\omega{}x)[g_i(x)f_f(x)+ \nonumber\\
&& g_f(x)f_i(x)]x^2 dx ,
\end{eqnarray*}
where $\kappa=(l-j)(2j+1)$, $l$ and $j$ are the quantum numbers of
the electron orbital and total angular momenta, $h_1^{(1)}$ is the
Hankel function of the first type.

We obtain
$\langle{}1s_{1/2}|{\emph{m}}_1^{M}|1s_{1/2}\rangle =
-2.22\times10^{-4}$, $E_{M1} \simeq 0.4$~eV
for the $^{229}$Th$^{89+}$ ion and
$\langle{}2s_{1/2}|{\emph{m}}_1^{M}|2s_{1/2}\rangle =
-3.66\times10^{-5}$, $E_{M1} \simeq 0.07$~eV
for the $^{229}$Th$^{87+}$ ion.

\begin{figure}
 \includegraphics[angle=0,width=0.8\hsize,keepaspectratio]{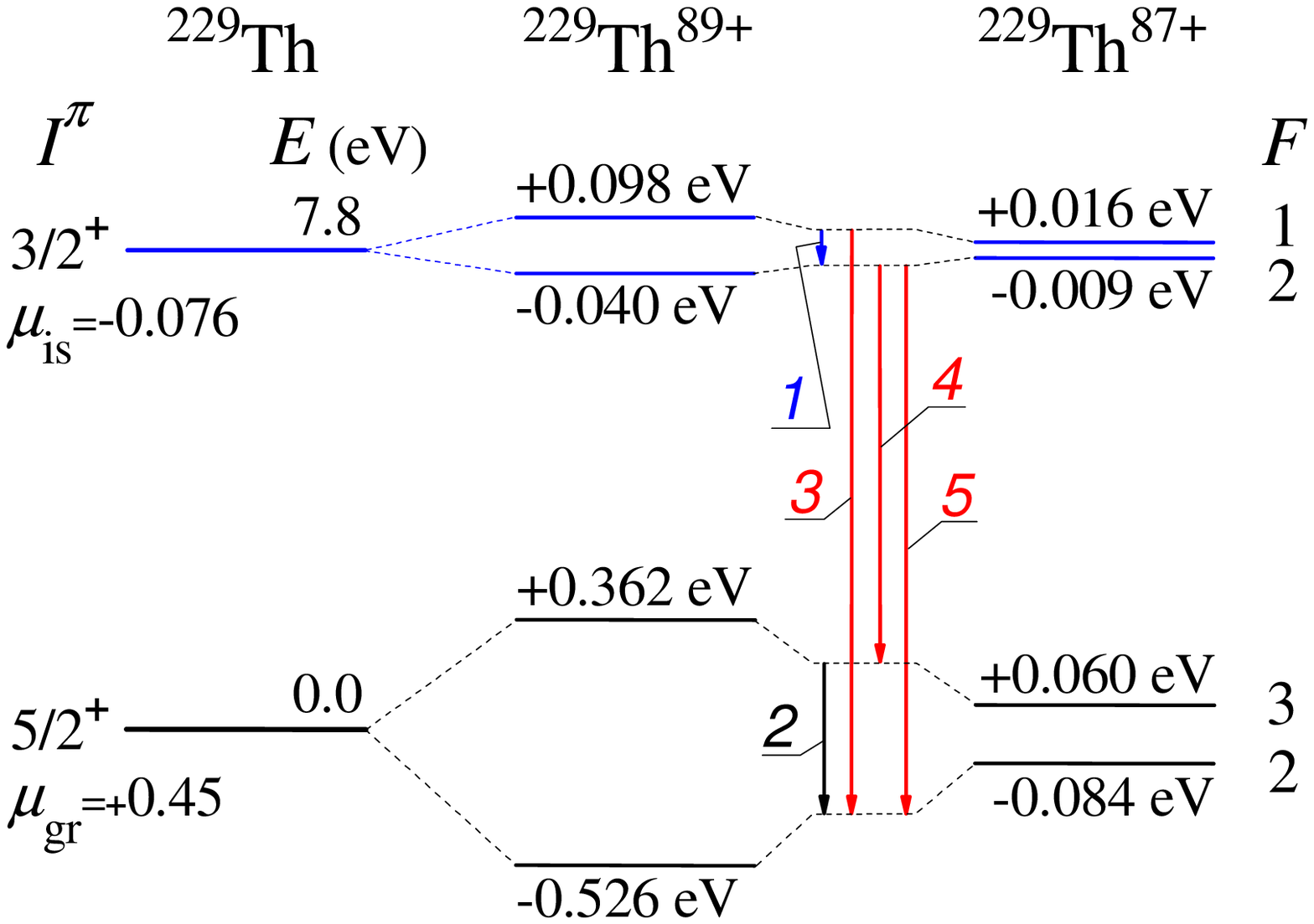}
 \caption{(color online). The magnetic hyperfine structure of the
 low energy doublet in $^{229}$Th$^{89+}$, $^{229}$Th$^{87+}$
 and the $M1$ transitions between sublevels. The sublevels  energy is counted
 from the energy of the level in the bare $^{229}$Th nucleus.}
  \label{fig:Levels-Transitions}
\end{figure}

The new energies of the $F=2$ sublevels are found as
\cite{Davydov-65}
\begin{eqnarray}
\lefteqn{ E_{|(4\mp1)/2^+,F=2\rangle{}' } =
\frac{E_{|3/2^+,F=2\rangle}+E_{|5/2^+,F=2\rangle}}{2} \pm
 }\nonumber\\
&&{} \frac{ \sqrt{(E_{|3/2^+,F=2\rangle}-E_{|5/2^+,F=2\rangle})^2
+ (2E_{M1})^2} }{2},
\end{eqnarray}
The positions of the sublevels are shown in
Fig.~\ref{fig:Levels-Transitions} and their energies are quoted in
Tables~\ref{tab:HPS_229Th89+} -- \ref{tab:HPS_229Th87+}. The
$^{229}$Th$^{89+}$ energies are in satisfactory correspondence
with the data of \cite{Karpeshin-98}. The MHF splitting for
$^{229}$Th$^{87+}$ shown in Fig.~\ref{fig:Levels-Transitions} are
calculated for the first time.

%
%  Figure 2
%

\section{Transitions between sublevels}

In this section we calculate the probability of spontaneous
transitions between the sublevels. According to
\cite{Berestetskii-80} the radiative decay width of the $M1$
transition can be written as
\begin{eqnarray}
 \Gamma_{rad}(M1)&=&\frac{4}{3}\omega^3
\frac{1}{2F_i+1} |\langle{}j_fI_fF_f \|\mu_{B}\hat{\sigma} + \nonumber\\
&& \mu_{gr(is)} \mu_N \hat{I}/I +
        \hat{\mu}_{tr} \mu_N) \| j_iI_iF_i \rangle |^2.
 \label{eq:Grad}
\end{eqnarray}
Here $\mu_{B}\hat{\sigma}$ is the operator of the electron
magnetic moment, $\mu_{gr(is)}\mu_N\hat{I}/I$ is the operator of
the nuclear transition between the hyperfine components of the
same level with spin $I$, and $\hat{\mu}_{tr}\mu_N$ is the
operator of the transition between the ground and isomeric states.

In the case $j_i=j_f=1/2$, $I_i=I_f=I_{gr(is)}$, for the
transition between the hyperfine components
$|F_i=I_{gr(is)}\pm1/2\rangle \rightarrow{}
|F_f=I_{gr(is)}\mp1/2\rangle$ from Eq.~(\ref{eq:Grad}) we obtain
\begin{eqnarray}
 \Gamma_{rad}(M1)&=& \frac{4}{3}e^2\frac{\omega^3}{m^2}
\frac{I_{gr(is)}+1/2\mp1/2}{2I_{gr(is)}+1}\times  \nonumber\\
&&{}
\left(1+\frac{m}{M_p}\frac{\mu_{gr(is)}}{2I_{gr(is)}}\right)^2.
 \label{eq:Grad1}
\end{eqnarray}
This expression coincides with $\Gamma_{rad}$ for the
$|F_i=I+1/2\rangle \rightarrow{} |F_f=I-1/2\rangle$ transition
quoted in \cite{Winston-63}. Clearly, the relative contribution
from the operator $\mu_{gr(is)}\mu_N\hat{I}/I$ given in the second
term in the parentheses of Eq.\~(\ref{eq:Grad1}), is small and
will be omitted below.

The probability associated with the operator $\hat{\mu}_{tr}\mu_N$
is negligibly small for both ions. This is a consequence of the
states mixing. The effect was already mentioned in
\cite{Karpeshin-98} for the $^{229}$Th$^{89+}$ ion. Although the
coefficient $b$ is several times smaller for the
$^{229}$Th$^{87+}$ ion ($b_{87+} = 0.0083$ versus $b_{89+} =
0.048$) it is nevertheless enough for enabling the noticeable
transitions between the hyperfine sublevels through the electron
spin-flip transition.

The calculated radiative decay widths $\Gamma_{rad}$ and the
associated transition times $\tau$ are given in
Table~\ref{tab:Grad}. Probabilities for the transitions 1 and 2
have been computed by means of Eq.~(\ref{eq:Grad1}). For the
transitions 3 and 4 the value from Eq.~(\ref{eq:Grad1}) was
further multiplied by $b^2$. The transition 5 involves the
electron spin-flip transitions $|I_i=F\mp1/2,F\rangle
\rightarrow{} |I_f=F\pm1/2,F\rangle$ (due to the mixing of the
$F=2$ states) and its radiative width is calculated according to
the equation
\begin{equation}
\Gamma_{rad}(M1) = \frac{4}{3}e^2\frac{\omega^3}{m^2}
\frac{b^2(1-b^2)}{4F(F+1)}  ,
 \label{eq:GradFF}
\end{equation}
which can be derived from (\ref{eq:Grad}).

The results for $^{229}$Th$^{89+}$ are in fair correspondence with
the data of \cite{Karpeshin-98}. (Notice that in
Ref.~\cite{Karpeshin-98}, as in other works up to 2007, it was
assumed that $E_{is}=3.5$ eV.)

%
% Table 3
%
\begin{table}
  \caption{Radiative widths $\Gamma_{rad}$ and times $\tau=\ln(2)/\Gamma_{rad}$
  for the transitions between the sublevels of MHF structure in
  $^{229}$Th$^{89+}$ and $^{229}$Th$^{87+}$.}
  \begin{tabular}{c|c|c|c|c|c}
    \hline
    \hline
\multicolumn{2}{c|}{Transition} & \multicolumn{2}{c|}{$^{229}$Th$^{89+}$} &\multicolumn{2}{c}{$^{229}$Th$^{87+}$} \\
                 \cline{1-6}
N & $I^{\pi}_i,F_i\rightarrow{}I^{\pi}_f,F_f$ & $\Gamma_{rad}$ (eV) & $\tau$  & $\Gamma_{rad}$ (eV) & $\tau$ \\
    \hline
1 & $3/2^+,1\rightarrow3/2^+,2$ & $6.2\times10^{-17}$ & 7.4 s & $3.8\times10^{-19}$ & 20 min \\
    \hline
2 & $5/2^+,3\rightarrow5/2^+,2$ & $1.1\times10^{-14}$ & 42 ms & $4.6\times10^{-17}$ & 10 s \\
       \hline
3 & $3/2^+,1\rightarrow5/2^+,2$ & $3.2\times10^{-14}$ & 14 ms & $8.0\times10^{-16}$ & 0.57 s \\
       \hline
4 & $3/2^+,2\rightarrow5/2^+,3$ & $2.2\times10^{-14}$ & 21 ms & $7.6\times10^{-16}$ & 0.61 s \\
       \hline
5 & $3/2^+,2\rightarrow5/2^+,2$ & $2.0\times10^{-15}$ & 0.23 s & $5.2\times10^{-17}$ & 8.7 s \\
       \hline
  \end{tabular}
  \label{tab:Grad}
 \end{table}

To demonstrate the increased probability of the transitions 3, 4
and 5, we compare their widths with the radiative width
$\Gamma_{rad}(3/2^+\rightarrow5/2^+)$ for the $M1$ isomeric
transition in the naked nucleus $^{229}$Th. In that case only the
operator $\hat{\mu}_{tr}\mu_N$ remains in Eq.~(\ref{eq:Grad}). We
express its matrix element in terms of the reduced probability of
the nuclear transition:
$$
B(M1;I_i\rightarrow{}I_f)= \frac{3}{4\pi} \frac{ |\langle{}I_f\|
\hat{\mu}_{tr} \|I_i\rangle|^2}{2I_i+1} .
$$
We then arrive at $B(M1;3/2^+\rightarrow{}5/2^+)=
(45/8\pi)\mu_N^2 B_{W.u.}(M1,3/2^+\rightarrow{}5/2^+) = 5.37\times
10^{-2}\mu_N^2$. (The factor $3/4\pi$ is accounted for by the
differences in the definition of the transition operators in the
atomic \cite{Berestetskii-80} and nuclear \cite{Bohr-98-II}
physics.) As a result we have
$\Gamma_{rad}(3/2^+\rightarrow5/2^+)= 3\times10^{-19}$ eV,
which corresponds to $\tau=26$ min.

\section{Conclusion}

In conclusion, the calculation of the magnetic hypefine structure
of the ground state doublet (the ground $5/2^+(0.0)$ and the low
energy $3/2^+(7.8$ eV) isomeric states) in the highly ionized ions
$^{229}$Th$^{89+}$ and $^{229}$Th$^{87+}$ have been performed.

We have demonstrated that in comparison with the point nucleus
model the Bohr-Weisskopf effect (finite distribution of nuclear
magnetization) decreases the hyperfine splitting by 3\% for the
ground state and by 6\% for the isomeric state. The energies of
the sublevels have been calculated by taking into account the
mixing of states with the same quantum number $F=2$. As a result
of mixing the energy difference between two $F$ states increases
by 0.04 eV in $^{229}$Th$^{89+}$ and 0.001 eV in
$^{229}$Th$^{87+}$.

The mixing of states (with the coefficients $b_{89+} = 0.048$ and
$b_{87+} = 0.0083$) has been taken into account in the estimation
of the probability of spontaneous transitions. We have found that
even in the $^{229}$Th$^{87+}$ ion the mixing leads to a large
increase of the probability of radiative transitions. The
transitions therefore are caused mainly by a small admixter of
other quantum states.

It is worth noting that the Bohr-Weisskopf effect completely
compensates the energy shift of the $|5/2^+,F=2\rangle$ level in
the $^{229}$Th$^{89+,\, 87+}$ ions, caused by the mixing. It also
has profound influence on the energy positions of the states with
$F=3$ and $F=1$. Therefore, the data on the hyperfine splittings
can be used for precise determination of the magnetic moments of
the nuclear $5/2^+(0.0)$ and $3/2^+(7.8$ eV) states.

Our findings can be useful for experiments with highly ionized
$^{229}$Th ions in the storage ring of accelerator complex.

\section{Acknowledgments}

One of the authors (E.T.) is grateful to Prof. S. Wycech, who has
drawn our attention to the importance of the calculation of the
MHF splitting in the $^{229}$Th$^{87+}$ ion for experimental
studies in the accelerator storage ring.

This research was supported by a grant of Russian Science
Foundation (project No 16-12-00001).

\bibstyle{apsrev}
%\bibliography{Bibliography_Th-229_ion}

\begin{thebibliography}{66}
\expandafter\ifx\csname
natexlab\endcsname\relax\def\natexlab#1{#1}\fi
\expandafter\ifx\csname bibnamefont\endcsname\relax
  \def\bibnamefont#1{#1}\fi
\expandafter\ifx\csname bibfnamefont\endcsname\relax
  \def\bibfnamefont#1{#1}\fi
\expandafter\ifx\csname citenamefont\endcsname\relax
  \def\citenamefont#1{#1}\fi
\expandafter\ifx\csname url\endcsname\relax
  \def\url#1{\texttt{#1}}\fi
\expandafter\ifx\csname
urlprefix\endcsname\relax\def\urlprefix{URL }\fi
\providecommand{\bibinfo}[2]{#2}
\providecommand{\eprint}[2][]{\url{#2}}

\bibitem[{\citenamefont{Beck et~al.}(2007)\citenamefont{Beck, Becker,
  Beiersdorfer, Brown, Moody, Wilhelmy, Porter, Kilbourne, and
  Kelley}}]{Beck-07}
\bibinfo{author}{\bibfnamefont{B.~R.} \bibnamefont{Beck}},
  \bibinfo{author}{\bibfnamefont{J.~A.} \bibnamefont{Becker}},
  \bibinfo{author}{\bibfnamefont{P.}~\bibnamefont{Beiersdorfer}},
  \bibinfo{author}{\bibfnamefont{G.~V.} \bibnamefont{Brown}},
  \bibinfo{author}{\bibfnamefont{K.~J.} \bibnamefont{Moody}},
  \bibinfo{author}{\bibfnamefont{J.~B.} \bibnamefont{Wilhelmy}},
  \bibinfo{author}{\bibfnamefont{F.~S.} \bibnamefont{Porter}},
  \bibinfo{author}{\bibfnamefont{C.~A.} \bibnamefont{Kilbourne}},
  \bibnamefont{and} \bibinfo{author}{\bibfnamefont{R.~L.}
  \bibnamefont{Kelley}}, \bibinfo{journal}{Phys. Rev. Lett.}
  \textbf{\bibinfo{volume}{98}}, \bibinfo{pages}{142501}
  (\bibinfo{year}{2007}).

\bibitem[{Bec()}]{Beck-R}
\bibinfo{note}{B. R. Beck, J. A. Becker, P. Beiersdorfer, G. V. Brown, K. J.
  Moody, J. B. Wilhelmy, F. S. Porter, C. A. Kilbourne and R. L. Kelley, Report
  LLNL-PROC-415170.}

\bibitem[{\citenamefont{Reich and Helmer}(1990)}]{Reich-90}
\bibinfo{author}{\bibfnamefont{C.~W.} \bibnamefont{Reich}} \bibnamefont{and}
  \bibinfo{author}{\bibfnamefont{R.~G.} \bibnamefont{Helmer}},
  \bibinfo{journal}{Phys. Rev. Lett.} \textbf{\bibinfo{volume}{64}},
  \bibinfo{pages}{271} (\bibinfo{year}{1990}).

\bibitem[{\citenamefont{Helmer and Reich}(1994)}]{Helmer-94}
\bibinfo{author}{\bibfnamefont{R.~G.} \bibnamefont{Helmer}} \bibnamefont{and}
  \bibinfo{author}{\bibfnamefont{C.~W.} \bibnamefont{Reich}},
  \bibinfo{journal}{Phys. Rev. C} \textbf{\bibinfo{volume}{49}},
  \bibinfo{pages}{1845} (\bibinfo{year}{1994}).

\bibitem[{\citenamefont{Irwin and Kim}(1997)}]{Irwin-97}
\bibinfo{author}{\bibfnamefont{G.~M.} \bibnamefont{Irwin}} \bibnamefont{and}
  \bibinfo{author}{\bibfnamefont{K.~H.} \bibnamefont{Kim}},
  \bibinfo{journal}{Phys. Rev. Lett.} \textbf{\bibinfo{volume}{79}},
  \bibinfo{pages}{990} (\bibinfo{year}{1997}).

\bibitem[{\citenamefont{Richardson et~al.}(1998)\citenamefont{Richardson,
  Benton, Evans, Griffith, and Tungate}}]{Richardson-98}
\bibinfo{author}{\bibfnamefont{D.~S.} \bibnamefont{Richardson}},
  \bibinfo{author}{\bibfnamefont{D.~M.} \bibnamefont{Benton}},
  \bibinfo{author}{\bibfnamefont{D.~E.} \bibnamefont{Evans}},
  \bibinfo{author}{\bibfnamefont{J.~A.~R.} \bibnamefont{Griffith}},
  \bibnamefont{and} \bibinfo{author}{\bibfnamefont{G.}~\bibnamefont{Tungate}},
  \bibinfo{journal}{Phys. Rev. Lett.} \textbf{\bibinfo{volume}{80}},
  \bibinfo{pages}{3206} (\bibinfo{year}{1998}).

\bibitem[{\citenamefont{Shaw et~al.}(1999)\citenamefont{Shaw, Young, Cooper,
  and Webb}}]{Shaw-99}
\bibinfo{author}{\bibfnamefont{R.~W.} \bibnamefont{Shaw}},
  \bibinfo{author}{\bibfnamefont{J.~P.} \bibnamefont{Young}},
  \bibinfo{author}{\bibfnamefont{S.~P.} \bibnamefont{Cooper}},
  \bibnamefont{and} \bibinfo{author}{\bibfnamefont{O.~F.} \bibnamefont{Webb}},
  \bibinfo{journal}{Phys. Rev. Lett.} \textbf{\bibinfo{volume}{82}},
  \bibinfo{pages}{1109} (\bibinfo{year}{1999}).

\bibitem[{\citenamefont{Utter et~al.}(1999)\citenamefont{Utter, Beiersdorfer,
  Barnes, Lougheed, Crespo Lopez-Urrutia, Becker, and Weiss}}]{Utter-99}
\bibinfo{author}{\bibfnamefont{S.~B.} \bibnamefont{Utter}},
  \bibinfo{author}{\bibfnamefont{P.}~\bibnamefont{Beiersdorfer}},
  \bibinfo{author}{\bibfnamefont{A.}~\bibnamefont{Barnes}},
  \bibinfo{author}{\bibfnamefont{R.~W.} \bibnamefont{Lougheed}},
  \bibinfo{author}{\bibfnamefont{J.~R.} \bibnamefont{Crespo Lopez-Urrutia}},
  \bibinfo{author}{\bibfnamefont{J.~A.} \bibnamefont{Becker}},
  \bibnamefont{and} \bibinfo{author}{\bibfnamefont{M.~S.} \bibnamefont{Weiss}},
  \bibinfo{journal}{Phys. Rev. Lett.} \textbf{\bibinfo{volume}{82}},
  \bibinfo{pages}{505} (\bibinfo{year}{1999}).

\bibitem[{Kik()}]{Kikunaga-05}
\bibinfo{note}{H. Kikunaga, Y. Kasamatsu, K. Takamiya, T. Mitsugashira, M.
  Hara, T. Ohtsuki, H. Yuki, A. Shinohara, S. Shibata, N. Kinoshita, A.
  Yokoyama, and T. Nakanishi. Research Rep. Lab. Nucl. Sci., Tohoku University,
  {\bf{38}}, 25 (2005)}.

\bibitem[{\citenamefont{Campbell et~al.}(2009)\citenamefont{Campbell, Steele,
  Churchill, DePalatis, Naylor, Matsukevich, Kuzmich, and
  Chapman}}]{Campbell-09}
\bibinfo{author}{\bibfnamefont{C.~J.} \bibnamefont{Campbell}},
  \bibinfo{author}{\bibfnamefont{A.~V.} \bibnamefont{Steele}},
  \bibinfo{author}{\bibfnamefont{L.~R.} \bibnamefont{Churchill}},
  \bibinfo{author}{\bibfnamefont{M.~V.} \bibnamefont{DePalatis}},
  \bibinfo{author}{\bibfnamefont{D.~E.} \bibnamefont{Naylor}},
  \bibinfo{author}{\bibfnamefont{D.~N.} \bibnamefont{Matsukevich}},
  \bibinfo{author}{\bibfnamefont{A.}~\bibnamefont{Kuzmich}}, \bibnamefont{and}
  \bibinfo{author}{\bibfnamefont{M.~S.} \bibnamefont{Chapman}},
  \bibinfo{journal}{Phys. Rev. Lett.} \textbf{\bibinfo{volume}{102}},
  \bibinfo{pages}{233004} (\bibinfo{year}{2009}).

\bibitem[{\citenamefont{Campbell et~al.}(2011)\citenamefont{Campbell, Radnaev,
  and Kuzmich}}]{Campbell-11}
\bibinfo{author}{\bibfnamefont{C.~J.} \bibnamefont{Campbell}},
  \bibinfo{author}{\bibfnamefont{A.~G.} \bibnamefont{Radnaev}},
  \bibnamefont{and} \bibinfo{author}{\bibfnamefont{A.}~\bibnamefont{Kuzmich}},
  \bibinfo{journal}{Phys. Rev. Lett.} \textbf{\bibinfo{volume}{106}},
  \bibinfo{pages}{223001} (\bibinfo{year}{2011}).

\bibitem[{\citenamefont{Campbell et~al.}(2012)\citenamefont{Campbell, Radnaev,
  Kuzmich, Dzuba, Flambaum, and Derevianko}}]{Campbell-12}
\bibinfo{author}{\bibfnamefont{C.~J.} \bibnamefont{Campbell}},
  \bibinfo{author}{\bibfnamefont{A.~G.} \bibnamefont{Radnaev}},
  \bibinfo{author}{\bibfnamefont{A.}~\bibnamefont{Kuzmich}},
  \bibinfo{author}{\bibfnamefont{V.~A.} \bibnamefont{Dzuba}},
  \bibinfo{author}{\bibfnamefont{V.~V.} \bibnamefont{Flambaum}},
  \bibnamefont{and}
  \bibinfo{author}{\bibfnamefont{A.}~\bibnamefont{Derevianko}},
  \bibinfo{journal}{Phys. Rev. Lett.} \textbf{\bibinfo{volume}{108}},
  \bibinfo{pages}{120802} (\bibinfo{year}{2012}).

\bibitem[{\citenamefont{Hehlen et~al.}(2013)\citenamefont{Hehlen, Greco,
  Rellergert, Sullivan, DeMille, Jackson, Hudson, and Torgerson}}]{Hehlen-13}
\bibinfo{author}{\bibfnamefont{M.~P.} \bibnamefont{Hehlen}},
  \bibinfo{author}{\bibfnamefont{R.~R.} \bibnamefont{Greco}},
  \bibinfo{author}{\bibfnamefont{W.~G.} \bibnamefont{Rellergert}},
  \bibinfo{author}{\bibfnamefont{S.~T.} \bibnamefont{Sullivan}},
  \bibinfo{author}{\bibfnamefont{D.}~\bibnamefont{DeMille}},
  \bibinfo{author}{\bibfnamefont{R.~A.} \bibnamefont{Jackson}},
  \bibinfo{author}{\bibfnamefont{E.~R.} \bibnamefont{Hudson}},
  \bibnamefont{and} \bibinfo{author}{\bibfnamefont{J.~R.}
  \bibnamefont{Torgerson}}, \bibinfo{journal}{J. Luminescence}
  \textbf{\bibinfo{volume}{49}}, \bibinfo{pages}{133} (\bibinfo{year}{2013}).

\bibitem[{\citenamefont{Jeet et~al.}(2015)\citenamefont{Jeet, Schneider,
  Sullivan, Rellergert, Mirzadeh, Cassanho, Jenssen, Tkalya, and
  Hudson}}]{Jeet-15}
\bibinfo{author}{\bibfnamefont{J.}~\bibnamefont{Jeet}},
  \bibinfo{author}{\bibfnamefont{C.}~\bibnamefont{Schneider}},
  \bibinfo{author}{\bibfnamefont{S.~T.} \bibnamefont{Sullivan}},
  \bibinfo{author}{\bibfnamefont{W.~G.} \bibnamefont{Rellergert}},
  \bibinfo{author}{\bibfnamefont{S.}~\bibnamefont{Mirzadeh}},
  \bibinfo{author}{\bibfnamefont{A.}~\bibnamefont{Cassanho}},
  \bibinfo{author}{\bibfnamefont{H.~P.} \bibnamefont{Jenssen}},
  \bibinfo{author}{\bibfnamefont{E.~V.} \bibnamefont{Tkalya}},
  \bibnamefont{and} \bibinfo{author}{\bibfnamefont{E.~R.}
  \bibnamefont{Hudson}}, \bibinfo{journal}{Phys. Rev. Lett.}
  \textbf{\bibinfo{volume}{114}}, \bibinfo{pages}{253001}
  (\bibinfo{year}{2015}).

\bibitem[{\citenamefont{Yamaguchi et~al.}(2015)\citenamefont{Yamaguchi, Kolbe,
  Kaser, Reichel, Gottwald, and Peik}}]{Yamaguchi-15}
\bibinfo{author}{\bibfnamefont{A.}~\bibnamefont{Yamaguchi}},
  \bibinfo{author}{\bibfnamefont{M.}~\bibnamefont{Kolbe}},
  \bibinfo{author}{\bibfnamefont{H.}~\bibnamefont{Kaser}},
  \bibinfo{author}{\bibfnamefont{T.}~\bibnamefont{Reichel}},
  \bibinfo{author}{\bibfnamefont{A.}~\bibnamefont{Gottwald}}, \bibnamefont{and}
  \bibinfo{author}{\bibfnamefont{E.}~\bibnamefont{Peik}}, \bibinfo{journal}{New
  J. Phys.} \textbf{\bibinfo{volume}{17}}, \bibinfo{pages}{053053}
  (\bibinfo{year}{2015}).

\bibitem[{\citenamefont{von~der Wense et~al.}(2016)\citenamefont{von~der Wense,
  Seiferle, Laatiaoui, Neumayr1, Maier, Wirth, Mokry, Runke, Eberhardt,
  Dullmann et~al.}}]{Wense-16}
\bibinfo{author}{\bibfnamefont{L.}~\bibnamefont{von~der Wense}},
  \bibinfo{author}{\bibfnamefont{B.}~\bibnamefont{Seiferle}},
  \bibinfo{author}{\bibfnamefont{M.}~\bibnamefont{Laatiaoui}},
  \bibinfo{author}{\bibfnamefont{J.~B.} \bibnamefont{Neumayr1}},
  \bibinfo{author}{\bibfnamefont{H.-J.} \bibnamefont{Maier}},
  \bibinfo{author}{\bibfnamefont{H.-F.} \bibnamefont{Wirth}},
  \bibinfo{author}{\bibfnamefont{C.}~\bibnamefont{Mokry}},
  \bibinfo{author}{\bibfnamefont{J.}~\bibnamefont{Runke}},
  \bibinfo{author}{\bibfnamefont{K.}~\bibnamefont{Eberhardt}},
  \bibinfo{author}{\bibfnamefont{C.~E.} \bibnamefont{Dullmann}},
  \bibnamefont{et~al.}, \bibinfo{journal}{Nature}
  \textbf{\bibinfo{volume}{533}}, \bibinfo{pages}{47} (\bibinfo{year}{2016}).

\bibitem[{\citenamefont{Strizhov and Tkalya}(1991)}]{Strizhov-91}
\bibinfo{author}{\bibfnamefont{V.~F.} \bibnamefont{Strizhov}} \bibnamefont{and}
  \bibinfo{author}{\bibfnamefont{E.~V.} \bibnamefont{Tkalya}},
  \bibinfo{journal}{Sov. Phys. JETP} \textbf{\bibinfo{volume}{72}},
  \bibinfo{pages}{387} (\bibinfo{year}{1991}).

\bibitem[{\citenamefont{Tkalya}(1992{\natexlab{a}})}]{Tkalya-92}
\bibinfo{author}{\bibfnamefont{E.~V.} \bibnamefont{Tkalya}},
  \bibinfo{journal}{Nucl. Phys. A} \textbf{\bibinfo{volume}{539}},
  \bibinfo{pages}{209} (\bibinfo{year}{1992}{\natexlab{a}}).

\bibitem[{\citenamefont{Wycech and Zylicz}(1993)}]{Wycech-93}
\bibinfo{author}{\bibfnamefont{S.}~\bibnamefont{Wycech}} \bibnamefont{and}
  \bibinfo{author}{\bibfnamefont{J.}~\bibnamefont{Zylicz}},
  \bibinfo{journal}{Acta Phys. Pol. B} \textbf{\bibinfo{volume}{24}},
  \bibinfo{pages}{637} (\bibinfo{year}{1993}).

\bibitem[{\citenamefont{Tkalya et~al.}(1996)\citenamefont{Tkalya, Varlamov,
  Lomonosov, and Nikulin}}]{Tkalya-96}
\bibinfo{author}{\bibfnamefont{E.~V.} \bibnamefont{Tkalya}},
  \bibinfo{author}{\bibfnamefont{V.~O.} \bibnamefont{Varlamov}},
  \bibinfo{author}{\bibfnamefont{V.~V.} \bibnamefont{Lomonosov}},
  \bibnamefont{and} \bibinfo{author}{\bibfnamefont{S.~A.}
  \bibnamefont{Nikulin}}, \bibinfo{journal}{Phys. Scr.}
  \textbf{\bibinfo{volume}{53}}, \bibinfo{pages}{296} (\bibinfo{year}{1996}).

\bibitem[{\citenamefont{Dykhne et~al.}(1996)\citenamefont{Dykhne, Eremin, and
  Tkalya}}]{Dykhne-96}
\bibinfo{author}{\bibfnamefont{A.~M.} \bibnamefont{Dykhne}},
  \bibinfo{author}{\bibfnamefont{N.~V.} \bibnamefont{Eremin}},
  \bibnamefont{and} \bibinfo{author}{\bibfnamefont{E.~V.}
  \bibnamefont{Tkalya}}, \bibinfo{journal}{JETP Lett.}
  \textbf{\bibinfo{volume}{64}}, \bibinfo{pages}{345} (\bibinfo{year}{1996}).

\bibitem[{\citenamefont{Karpeshin et~al.}(1998)\citenamefont{Karpeshin, Wycech,
  Band, Trzhaskovskaya, Pfutzner, and Zylicz}}]{Karpeshin-98}
\bibinfo{author}{\bibfnamefont{F.~F.} \bibnamefont{Karpeshin}},
  \bibinfo{author}{\bibfnamefont{S.}~\bibnamefont{Wycech}},
  \bibinfo{author}{\bibfnamefont{I.~M.} \bibnamefont{Band}},
  \bibinfo{author}{\bibfnamefont{M.~B.} \bibnamefont{Trzhaskovskaya}},
  \bibinfo{author}{\bibfnamefont{M.}~\bibnamefont{Pfutzner}}, \bibnamefont{and}
  \bibinfo{author}{\bibfnamefont{J.}~\bibnamefont{Zylicz}},
  \bibinfo{journal}{Phys. Rev. C} \textbf{\bibinfo{volume}{57}},
  \bibinfo{pages}{3085} (\bibinfo{year}{1998}).

\bibitem[{\citenamefont{Dykhne and Tkalya}(1998{\natexlab{a}})}]{Dykhne-98_ME}
\bibinfo{author}{\bibfnamefont{A.~M.} \bibnamefont{Dykhne}} \bibnamefont{and}
  \bibinfo{author}{\bibfnamefont{E.~V.} \bibnamefont{Tkalya}},
  \bibinfo{journal}{JETP Lett.} \textbf{\bibinfo{volume}{67}},
  \bibinfo{pages}{251} (\bibinfo{year}{1998}{\natexlab{a}}).

\bibitem[{\citenamefont{Dykhne and Tkalya}(1998{\natexlab{b}})}]{Dykhne-98}
\bibinfo{author}{\bibfnamefont{A.~M.} \bibnamefont{Dykhne}} \bibnamefont{and}
  \bibinfo{author}{\bibfnamefont{E.~V.} \bibnamefont{Tkalya}},
  \bibinfo{journal}{JETP Lett.} \textbf{\bibinfo{volume}{67}},
  \bibinfo{pages}{549} (\bibinfo{year}{1998}{\natexlab{b}}).

\bibitem[{\citenamefont{Tkalya}(2000)}]{Tkalya-00-JETPL}
\bibinfo{author}{\bibfnamefont{E.~V.} \bibnamefont{Tkalya}},
  \bibinfo{journal}{JETP Lett.} \textbf{\bibinfo{volume}{71}},
  \bibinfo{pages}{311} (\bibinfo{year}{2000}).

\bibitem[{\citenamefont{Tkalya et~al.}(2000)\citenamefont{Tkalya, Zherikhin,
  and Zhudov}}]{Tkalya-00-PRC}
\bibinfo{author}{\bibfnamefont{E.~V.} \bibnamefont{Tkalya}},
  \bibinfo{author}{\bibfnamefont{A.~N.} \bibnamefont{Zherikhin}},
  \bibnamefont{and} \bibinfo{author}{\bibfnamefont{V.~I.}
  \bibnamefont{Zhudov}}, \bibinfo{journal}{Phys. Rev. C}
  \textbf{\bibinfo{volume}{61}}, \bibinfo{pages}{064308}
  (\bibinfo{year}{2000}).

\bibitem[{\citenamefont{Tkalya}(2003)}]{Tkalya-03}
\bibinfo{author}{\bibfnamefont{E.~V.} \bibnamefont{Tkalya}},
  \bibinfo{journal}{Physics-Uspekhi} \textbf{\bibinfo{volume}{46}},
  \bibinfo{pages}{315} (\bibinfo{year}{2003}).

\bibitem[{\citenamefont{Flambaum}(2006)}]{Flambaum-06}
\bibinfo{author}{\bibfnamefont{V.~V.} \bibnamefont{Flambaum}},
  \bibinfo{journal}{Phys. Rev. Lett.} \textbf{\bibinfo{volume}{97}},
  \bibinfo{pages}{092502} (\bibinfo{year}{2006}).

\bibitem[{\citenamefont{Berengut et~al.}(2009)\citenamefont{Berengut, Dzuba,
  Flambaum, and Porsev}}]{Berengut-09}
\bibinfo{author}{\bibfnamefont{J.~C.} \bibnamefont{Berengut}},
  \bibinfo{author}{\bibfnamefont{V.~A.} \bibnamefont{Dzuba}},
  \bibinfo{author}{\bibfnamefont{V.~V.} \bibnamefont{Flambaum}},
  \bibnamefont{and} \bibinfo{author}{\bibfnamefont{S.~G.}
  \bibnamefont{Porsev}}, \bibinfo{journal}{Phys. Rev. Lett.}
  \textbf{\bibinfo{volume}{102}}, \bibinfo{pages}{210801}
  (\bibinfo{year}{2009}).

\bibitem[{\citenamefont{Litvinova et~al.}(2009)\citenamefont{Litvinova,
  Feldmeier, Dobaczewski, and Flambaum}}]{Litvinova-09}
\bibinfo{author}{\bibfnamefont{E.}~\bibnamefont{Litvinova}},
  \bibinfo{author}{\bibfnamefont{H.}~\bibnamefont{Feldmeier}},
  \bibinfo{author}{\bibfnamefont{J.}~\bibnamefont{Dobaczewski}},
  \bibnamefont{and} \bibinfo{author}{\bibfnamefont{V.}~\bibnamefont{Flambaum}},
  \bibinfo{journal}{Phys. Rev. C} \textbf{\bibinfo{volume}{79}},
  \bibinfo{pages}{064303} (\bibinfo{year}{2009}).

\bibitem[{\citenamefont{Tkalya}(2011)}]{Tkalya-11}
\bibinfo{author}{\bibfnamefont{E.~V.} \bibnamefont{Tkalya}},
  \bibinfo{journal}{Phys. Rev. Lett.} \textbf{\bibinfo{volume}{106}},
  \bibinfo{pages}{162501} (\bibinfo{year}{2011}).

\bibitem[{\citenamefont{Kazakov et~al.}(2012)\citenamefont{Kazakov, Litvinov,
  Romanenko, Yatsenko, Romanenko, Schreitl, Winkler, and Schumm}}]{Kazakov-12}
\bibinfo{author}{\bibfnamefont{G.~A.} \bibnamefont{Kazakov}},
  \bibinfo{author}{\bibfnamefont{A.~N.} \bibnamefont{Litvinov}},
  \bibinfo{author}{\bibfnamefont{V.~I.} \bibnamefont{Romanenko}},
  \bibinfo{author}{\bibfnamefont{L.~P.} \bibnamefont{Yatsenko}},
  \bibinfo{author}{\bibfnamefont{A.~V.} \bibnamefont{Romanenko}},
  \bibinfo{author}{\bibfnamefont{M.}~\bibnamefont{Schreitl}},
  \bibinfo{author}{\bibfnamefont{G.}~\bibnamefont{Winkler}}, \bibnamefont{and}
  \bibinfo{author}{\bibfnamefont{T.}~\bibnamefont{Schumm}},
  \bibinfo{journal}{New J. Phys.} \textbf{\bibinfo{volume}{14}},
  \bibinfo{pages}{083019} (\bibinfo{year}{2012}).

\bibitem[{\citenamefont{Beloy}(2014)}]{Beloy-14}
\bibinfo{author}{\bibfnamefont{K.}~\bibnamefont{Beloy}},
  \bibinfo{journal}{Phys. Rev. Lett.} \textbf{\bibinfo{volume}{112}},
  \bibinfo{pages}{062503} (\bibinfo{year}{2014}).

\bibitem[{\citenamefont{Tkalya et~al.}(2015)\citenamefont{Tkalya, Schneider,
  Jeet, and Hudson}}]{Tkalya-15}
\bibinfo{author}{\bibfnamefont{E.~V.} \bibnamefont{Tkalya}},
  \bibinfo{author}{\bibfnamefont{C.}~\bibnamefont{Schneider}},
  \bibinfo{author}{\bibfnamefont{J.}~\bibnamefont{Jeet}}, \bibnamefont{and}
  \bibinfo{author}{\bibfnamefont{E.~R.} \bibnamefont{Hudson}},
  \bibinfo{journal}{Phys. Rev. C} \textbf{\bibinfo{volume}{92}},
  \bibinfo{pages}{054324} (\bibinfo{year}{2015}).

\bibitem[{\citenamefont{Flambaum and Wiringa}(2009)}]{Flambaum-09}
\bibinfo{author}{\bibfnamefont{V.~V.} \bibnamefont{Flambaum}} \bibnamefont{and}
  \bibinfo{author}{\bibfnamefont{R.~B.} \bibnamefont{Wiringa}},
  \bibinfo{journal}{Phys. Rev. C} \textbf{\bibinfo{volume}{79}},
  \bibinfo{pages}{034302} (\bibinfo{year}{2009}).

\bibitem[{\citenamefont{Dzuba and Flambaum}(2010)}]{Dzuba-10}
\bibinfo{author}{\bibfnamefont{V.~A.} \bibnamefont{Dzuba}} \bibnamefont{and}
  \bibinfo{author}{\bibfnamefont{V.~V.} \bibnamefont{Flambaum}},
  \bibinfo{journal}{Phys. Rev. A} \textbf{\bibinfo{volume}{81}},
  \bibinfo{pages}{034501} (\bibinfo{year}{2010}).

\bibitem[{\citenamefont{Skripnikov et~al.}(2014)\citenamefont{Skripnikov,
  Petrov, Titov, and Flambaum}}]{Skripnikov-14}
\bibinfo{author}{\bibfnamefont{L.~V.} \bibnamefont{Skripnikov}},
  \bibinfo{author}{\bibfnamefont{A.~N.} \bibnamefont{Petrov}},
  \bibinfo{author}{\bibfnamefont{A.~V.} \bibnamefont{Titov}}, \bibnamefont{and}
  \bibinfo{author}{\bibfnamefont{V.~V.} \bibnamefont{Flambaum}},
  \bibinfo{journal}{Phys. Rev. Lett.} \textbf{\bibinfo{volume}{113}},
  \bibinfo{pages}{263006} (\bibinfo{year}{2014}).

\bibitem[{Fla()}]{Flambaum-16}
\bibinfo{note}{V.V. Flambaum, arXiv:1603.05753v1 (2016)}.

\bibitem[{\citenamefont{Porsev and Flambaum}(2010)}]{Porsev-10-PRA}
\bibinfo{author}{\bibfnamefont{S.~G.} \bibnamefont{Porsev}} \bibnamefont{and}
  \bibinfo{author}{\bibfnamefont{V.~V.} \bibnamefont{Flambaum}},
  \bibinfo{journal}{Phys. Rev. A} \textbf{\bibinfo{volume}{81}},
  \bibinfo{pages}{042516} (\bibinfo{year}{2010}).

\bibitem[{\citenamefont{Tkalya}(1992{\natexlab{b}})}]{Tkalya-92-JETPL}
\bibinfo{author}{\bibfnamefont{E.~V.} \bibnamefont{Tkalya}},
  \bibinfo{journal}{JETP Lett.} \textbf{\bibinfo{volume}{55}},
  \bibinfo{pages}{211} (\bibinfo{year}{1992}{\natexlab{b}}).

\bibitem[{\citenamefont{Porsev et~al.}(2010)\citenamefont{Porsev, Flambaum,
  Peik, and Tamm}}]{Porsev-10-PRL}
\bibinfo{author}{\bibfnamefont{S.~G.} \bibnamefont{Porsev}},
  \bibinfo{author}{\bibfnamefont{V.~V.} \bibnamefont{Flambaum}},
  \bibinfo{author}{\bibfnamefont{E.}~\bibnamefont{Peik}}, \bibnamefont{and}
  \bibinfo{author}{\bibfnamefont{C.}~\bibnamefont{Tamm}},
  \bibinfo{journal}{Phys. Rev. Lett.} \textbf{\bibinfo{volume}{105}},
  \bibinfo{pages}{182501} (\bibinfo{year}{2010}).

\bibitem[{\citenamefont{Peik and Tamm}(2000)}]{Peik-03}
\bibinfo{author}{\bibfnamefont{E.}~\bibnamefont{Peik}} \bibnamefont{and}
  \bibinfo{author}{\bibfnamefont{C.}~\bibnamefont{Tamm}},
  \bibinfo{journal}{Europhys. Lett.} \textbf{\bibinfo{volume}{61}},
  \bibinfo{pages}{181} (\bibinfo{year}{2000}).

\bibitem[{\citenamefont{Rellergert et~al.}(2010)\citenamefont{Rellergert,
  DeMille, Greco, Hehlen, Torgerson, and Hudson}}]{Rellergert-10}
\bibinfo{author}{\bibfnamefont{W.~G.} \bibnamefont{Rellergert}},
  \bibinfo{author}{\bibfnamefont{D.}~\bibnamefont{DeMille}},
  \bibinfo{author}{\bibfnamefont{R.~R.} \bibnamefont{Greco}},
  \bibinfo{author}{\bibfnamefont{M.~P.} \bibnamefont{Hehlen}},
  \bibinfo{author}{\bibfnamefont{J.~R.} \bibnamefont{Torgerson}},
  \bibnamefont{and} \bibinfo{author}{\bibfnamefont{E.~R.}
  \bibnamefont{Hudson}}, \bibinfo{journal}{Phys. Rev. Lett.}
  \textbf{\bibinfo{volume}{104}}, \bibinfo{pages}{200802}
  (\bibinfo{year}{2010}).

\bibitem[{\citenamefont{Peik and Okhapkin}(2015)}]{Peik-15}
\bibinfo{author}{\bibfnamefont{E.}~\bibnamefont{Peik}} \bibnamefont{and}
  \bibinfo{author}{\bibfnamefont{M.}~\bibnamefont{Okhapkin}},
  \bibinfo{journal}{C. R. Phys.} \textbf{\bibinfo{volume}{16}},
  \bibinfo{pages}{516} (\bibinfo{year}{2015}).

\bibitem[{\citenamefont{Tkalya and Yatsenko}(2013)}]{Tkalya-13}
\bibinfo{author}{\bibfnamefont{E.~V.} \bibnamefont{Tkalya}} \bibnamefont{and}
  \bibinfo{author}{\bibfnamefont{L.~P.} \bibnamefont{Yatsenko}},
  \bibinfo{journal}{Laser Phys. Lett.} \textbf{\bibinfo{volume}{10}},
  \bibinfo{pages}{105808} (\bibinfo{year}{2013}).

\bibitem[{\citenamefont{Geissel et~al.}(1992)\citenamefont{Geissel, Beckert,
  Bosch et~al.}}]{Geissel-92}
\bibinfo{author}{\bibfnamefont{H.}~\bibnamefont{Geissel}},
  \bibinfo{author}{\bibfnamefont{K.}~\bibnamefont{Beckert}},
  \bibinfo{author}{\bibfnamefont{F.}~\bibnamefont{Bosch}},
  \bibnamefont{et~al.}, \bibinfo{journal}{Phys. Rev. Lett.}
  \textbf{\bibinfo{volume}{68}}, \bibinfo{pages}{3412} (\bibinfo{year}{1992}).

\bibitem[{\citenamefont{Klaft et~al.}(1994)\citenamefont{Klaft, Borneis, Engel
  et~al.}}]{Klaf-94}
\bibinfo{author}{\bibfnamefont{I.}~\bibnamefont{Klaft}},
  \bibinfo{author}{\bibfnamefont{S.}~\bibnamefont{Borneis}},
  \bibinfo{author}{\bibfnamefont{T.}~\bibnamefont{Engel}},
  \bibnamefont{et~al.}, \bibinfo{journal}{Phys. Lett. Lett.}
  \textbf{\bibinfo{volume}{73}}, \bibinfo{pages}{2425} (\bibinfo{year}{1994}).

\bibitem[{\citenamefont{Radon et~al.}(1997)\citenamefont{Radon, Kerscher,
  Schlitt et~al.}}]{Radon-97}
\bibinfo{author}{\bibfnamefont{T.}~\bibnamefont{Radon}},
  \bibinfo{author}{\bibfnamefont{T.}~\bibnamefont{Kerscher}},
  \bibinfo{author}{\bibfnamefont{B.}~\bibnamefont{Schlitt}},
  \bibnamefont{et~al.}, \bibinfo{journal}{Phys. Rev. Lett.}
  \textbf{\bibinfo{volume}{78}}, \bibinfo{pages}{4701} (\bibinfo{year}{1997}).

\bibitem[{\citenamefont{Ma et~al.}(2015)\citenamefont{Ma, Wen, Huang, Wang,
  Yuan, Wang, Sun, Mao, Yang, Xu et~al.}}]{Ma-15}
\bibinfo{author}{\bibfnamefont{X.}~\bibnamefont{Ma}},
  \bibinfo{author}{\bibfnamefont{W.~Q.} \bibnamefont{Wen}},
  \bibinfo{author}{\bibfnamefont{Z.~K.} \bibnamefont{Huang}},
  \bibinfo{author}{\bibfnamefont{H.~B.} \bibnamefont{Wang}},
  \bibinfo{author}{\bibfnamefont{Y.~J.} \bibnamefont{Yuan}},
  \bibinfo{author}{\bibfnamefont{M.}~\bibnamefont{Wang}},
  \bibinfo{author}{\bibfnamefont{Z.~Y.} \bibnamefont{Sun}},
  \bibinfo{author}{\bibfnamefont{L.~J.} \bibnamefont{Mao}},
  \bibinfo{author}{\bibfnamefont{J.~C.} \bibnamefont{Yang}},
  \bibinfo{author}{\bibfnamefont{H.~S.} \bibnamefont{Xu}},
  \bibnamefont{et~al.}, \bibinfo{journal}{Phys. Scr.}
  \textbf{\bibinfo{volume}{T166}}, \bibinfo{pages}{014012}
  (\bibinfo{year}{2015}).

\bibitem[{\citenamefont{Abragam}(1961)}]{Abragam-61}
\bibinfo{author}{\bibfnamefont{A.}~\bibnamefont{Abragam}},
  \emph{\bibinfo{title}{The Principles of Nuclear Magnetism}}
  (\bibinfo{publisher}{Clarendon Press}, \bibinfo{address}{Oxford, England},
  \bibinfo{year}{1961}).

\bibitem[{Nik()}]{Nikolaev-15}
\bibinfo{note}{A. V. Nikolaev, D. Lamoen and B. Partoens, arXiv:1503.05827v2
  (2015)}.

\bibitem[{FLA()}]{FLAPW}
\bibinfo{note}{A. V. Nikolaev, The FLAPW-Moscow code [registration number
  2015616990 (Russia) from 26/06/2015]}.

\bibitem[{\citenamefont{Dirac}(1930)}]{Dirac-30}
\bibinfo{author}{\bibfnamefont{P.~A.~M.} \bibnamefont{Dirac}},
  \bibinfo{journal}{Proc. Camb. Philos. Soc.} \textbf{\bibinfo{volume}{26}},
  \bibinfo{pages}{376} (\bibinfo{year}{1930}).

\bibitem[{\citenamefont{Perdew and Wang}(1992)}]{Perdew-92}
\bibinfo{author}{\bibfnamefont{J.~P.} \bibnamefont{Perdew}} \bibnamefont{and}
  \bibinfo{author}{\bibfnamefont{Y.}~\bibnamefont{Wang}},
  \bibinfo{journal}{Phys. Rev. B} \textbf{\bibinfo{volume}{45}},
  \bibinfo{pages}{13244} (\bibinfo{year}{1992}).

\bibitem[{\citenamefont{Perdew et~al.}(1996)\citenamefont{Perdew, Burke, and
  Ernzerhof}}]{Perdew-96}
\bibinfo{author}{\bibfnamefont{J.~P.} \bibnamefont{Perdew}},
  \bibinfo{author}{\bibfnamefont{K.}~\bibnamefont{Burke}}, \bibnamefont{and}
  \bibinfo{author}{\bibfnamefont{M.}~\bibnamefont{Ernzerhof}},
  \bibinfo{journal}{Phys. Rev. Lett.} \textbf{\bibinfo{volume}{77}},
  \bibinfo{pages}{3865} (\bibinfo{year}{1996}).

\bibitem[{\citenamefont{Perdew et~al.}(1997)\citenamefont{Perdew, Burke, and
  Ernzerho}}]{Perdew-97}
\bibinfo{author}{\bibfnamefont{J.~P.} \bibnamefont{Perdew}},
  \bibinfo{author}{\bibfnamefont{K.}~\bibnamefont{Burke}}, \bibnamefont{and}
  \bibinfo{author}{\bibfnamefont{M.}~\bibnamefont{Ernzerho}},
  \bibinfo{journal}{Phys. Rev. Lett.} \textbf{\bibinfo{volume}{78}},
  \bibinfo{pages}{1396} (\bibinfo{year}{1997}).

\bibitem[{\citenamefont{Bohr and Weisskopf}(1950)}]{Bohr-50}
\bibinfo{author}{\bibfnamefont{A.}~\bibnamefont{Bohr}} \bibnamefont{and}
  \bibinfo{author}{\bibfnamefont{V.~F.} \bibnamefont{Weisskopf}},
  \bibinfo{journal}{Phys. Rev.} \textbf{\bibinfo{volume}{77}},
  \bibinfo{pages}{94} (\bibinfo{year}{1950}).

\bibitem[{\citenamefont{Bellac}(1963)}]{LeBellac-63}
\bibinfo{author}{\bibfnamefont{M.~Le} \bibnamefont{Bellac}},
  \bibinfo{journal}{Nucl. Phys.} \textbf{\bibinfo{volume}{40}},
  \bibinfo{pages}{645} (\bibinfo{year}{1963}).

\bibitem[{\citenamefont{Bohr and Mottelson}(1998{\natexlab{a}})}]{Bohr-98-II}
\bibinfo{author}{\bibfnamefont{A.}~\bibnamefont{Bohr}} \bibnamefont{and}
  \bibinfo{author}{\bibfnamefont{B.~R.} \bibnamefont{Mottelson}},
  \emph{\bibinfo{title}{Nuclear Structure. Vol. II: Nuclear Deformations.}}
  (\bibinfo{publisher}{World Scientific}, \bibinfo{address}{London},
  \bibinfo{year}{1998}{\natexlab{a}}).

\bibitem[{\citenamefont{Varshalovich et~al.}(1988)\citenamefont{Varshalovich,
  Moskalev, and Khersonslii}}]{Varshalovich-88}
\bibinfo{author}{\bibfnamefont{D.~A.} \bibnamefont{Varshalovich}},
  \bibinfo{author}{\bibfnamefont{A.~N.} \bibnamefont{Moskalev}},
  \bibnamefont{and} \bibinfo{author}{\bibfnamefont{V.~K.}
  \bibnamefont{Khersonslii}}, \emph{\bibinfo{title}{Quantum Theory of Angular
  Momentum}} (\bibinfo{publisher}{World Scientific Publ.},
  \bibinfo{address}{London}, \bibinfo{year}{1988}).

\bibitem[{\citenamefont{Abramowitz and Stegun}(1964)}]{Abramowitz-64}
\bibinfo{author}{\bibfnamefont{M.}~\bibnamefont{Abramowitz}} \bibnamefont{and}
  \bibinfo{author}{\bibfnamefont{I.~A.} \bibnamefont{Stegun}},
  \emph{\bibinfo{title}{Handbook of Mathematical Functions}}
  (\bibinfo{publisher}{National Bureau of Standards},
  \bibinfo{address}{Washington, D.C.}, \bibinfo{year}{1964}).

\bibitem[{\citenamefont{Davydov}(1965)}]{Davydov-65}
\bibinfo{author}{\bibfnamefont{A.~S.} \bibnamefont{Davydov}},
  \emph{\bibinfo{title}{Quantum Mechanics}} (\bibinfo{publisher}{Pergamon
  Press}, \bibinfo{address}{Oxford, England}, \bibinfo{year}{1965}).

\bibitem[{\citenamefont{Bohr and Mottelson}(1998{\natexlab{b}})}]{Bohr-98-I}
\bibinfo{author}{\bibfnamefont{A.}~\bibnamefont{Bohr}} \bibnamefont{and}
  \bibinfo{author}{\bibfnamefont{B.~R.} \bibnamefont{Mottelson}},
  \emph{\bibinfo{title}{Nuclear Structure. Vol. I: Single-Particle Motion.}}
  (\bibinfo{publisher}{World Scientific}, \bibinfo{address}{London},
  \bibinfo{year}{1998}{\natexlab{b}}).

\bibitem[{\citenamefont{Blatt and Weisskopf}(1952)}]{Blatt-Weisskopf}
\bibinfo{author}{\bibfnamefont{J.~M.} \bibnamefont{Blatt}} \bibnamefont{and}
  \bibinfo{author}{\bibfnamefont{V.~F.} \bibnamefont{Weisskopf}},
  \emph{\bibinfo{title}{Theoretical Nuclear Physics}} (\bibinfo{publisher}{lohn
  Wiley and Sons, Inc.,}, \bibinfo{address}{New York}, \bibinfo{year}{1952}).

\bibitem[{\citenamefont{Berestetskii et~al.}(1982)\citenamefont{Berestetskii,
  Lifschitz, and Pitaevskii}}]{Berestetskii-80}
\bibinfo{author}{\bibfnamefont{V.~B.} \bibnamefont{Berestetskii}},
  \bibinfo{author}{\bibfnamefont{E.~M.} \bibnamefont{Lifschitz}},
  \bibnamefont{and} \bibinfo{author}{\bibfnamefont{L.~P.}
  \bibnamefont{Pitaevskii}}, \emph{\bibinfo{title}{Quantum Electrodynamics}}
  (\bibinfo{publisher}{Pergamon Press}, \bibinfo{address}{Oxford, England},
  \bibinfo{year}{1982}).

\bibitem[{\citenamefont{Winston}(1963)}]{Winston-63}
\bibinfo{author}{\bibfnamefont{R.}~\bibnamefont{Winston}},
  \bibinfo{journal}{Phys. Rev.} \textbf{\bibinfo{volume}{129}},
  \bibinfo{pages}{2766} (\bibinfo{year}{1963}).

\end{thebibliography}

\end{document}